\documentclass[12pt]{iopart}
\usepackage{graphicx,color}

\begin{document}


\title[Experimental effects of disorder cross-correlations]{Transmission in waveguides with compositional and structural disorder: experimental effects of disorder cross-correlations}

\author{O Dietz$^1$, U Kuhl$^{1,2}$, J.C. Hern\'{a}ndez-Herrej\'{o}n$^3$, and L. Tessieri$^{3,4}$}
\address{$^1$ Fachbereich Physik, Philipps-Universit\"{a}t Marburg, Renthof 5, D-35032 Marburg, Germany}
\address{$^2$ Laboratoire de Physique Mati{\`{e}}re Condens{\'{e}}e (LPMC/CNRS UMR7336), Universit{\'{e}} Nice-Sophia Antipolis, Parc Valrose, F-06108 Nice Cedex 2, France}
\address{$^3$ Instituto de F\'{\i}sica y Matem\'{a}ticas, Universidad Michoacana de San Nicol\'{a}s de Hidalgo, 58060, Morelia, Mexico}
\address{$^4$ Instituto dei Sistemi Complessi, via Madonna del Piano, 10; I-50019 Sesto Fiorentino, Italy}

\eads{\mailto{otto.dietz@physik.hu-berlin.de},\mailto{ulrich.kuhl@unice.fr}, \mailto{juliocesar@ifm.umich.mx}, \mailto{luca.tessieri@gmail.com}}

\date{31st August 2011}

\begin{abstract}
We analyse the single-mode transmission of microwaves in a guide with internal random structure. The waveguide contains scatterers characterised by random heights and positions, corresponding to compositional and structural disorder. We measure the effects of cross-correlations between two kinds of disorder, showing how they enhance or attenuate the experimentally found transmission gaps generated by long-range self-correlations. The results agree with the theoretical predictions obtained for the aperiodic Kronig-Penney model and prove that self- and cross-correlations have relevant effects also in finite disordered samples of small size.
\end{abstract}


\submitto{\NJP}

\maketitle


\section{Introduction}

At the end of the 1990s it was realised that, contrary to previously held beliefs, effective localisation-delocalisation transitions can take place in one-dimensional (1D) and quasi-1D random models, provided that disorder is endowed with specific long-range correlations~\cite{mou98,mou00,izr99a,izr99b}. In particular, the study of 1D models with correlated random potentials was greatly spurred by the seminal paper of Izrailev and Krokhin~\cite{izr99a,izr99b}, who derived an analytical formula for the localisation length in one-dimensional models with weak correlated disorder. The discovery triggered an intense research activity in the field of low-dimensional models, which are now understood to possess a much richer behaviour than once thought (for a review, see~\cite{mar08}).

The theoretical predictions about the existence of delocalised states in 1D or quasi-1D models with correlated disorder~\cite{dun90,mou98,mou00,izr99a,izr99b} soon found experimental confirmations in semiconductor superlattices~\cite{bel99} and microwave waveguides~\cite{kuh00a,kro02}. Since these early works, 1D models with correlated disorder have found applications in an increasing number of physical fields, including Bose-Einstein condensates~\cite{san07,lug09,bou09}, bilayered media~\cite{izr09}, structures with corrugated surfaces~\cite{izr03d,izr05c}, and DNA modelling~\cite{kro09,alb06}.

In most of these works, the system under consideration has a single self-correlated random potential. Recently, however, some theoretical studies have begun to explore more complex models, characterised by the simultaneous presence of two kinds of disorder. For instance, in~\cite{izr09} the authors considered the transport properties of generic bilayered structures formed by infinite arrays of alternating slabs of two different kinds. Since slabs of both types can have random widths, each structure is characterised not by one but by two random sequences which, in addition to self-correlations, can exhibit cross-correlations. Another model exhibiting two types of disorder is the aperiodic Kronig-Penney model with both compositional and structural disorder~\cite{her08,her10b,her10c}. For this model, it was possible to work out a perturbative expression for the localisation length of the quantum states valid for weak disorder with any kind of self- and cross-correlations~\cite{her08}. The theoretical results for the localisation length of electronic states showed that the interplay of structural and compositional disorder could enhance the delocalisation-localisation transition created by long-range self-correlations of each type of disorder. It was also shown that cross-correlations could affect the transmission properties of a long random sample, increasing or decreasing the transmittivity in the opaque windows corresponding to the localised states.

These findings posed the question of whether these effects could be experimentally observed in appropriately designed random microwave waveguides. In fact, the structure of the electronic states in a Kronig-Penney model and the transmission of the waveguide with point-like scatterers are closely related problems because the Schr\"{o}dinger equation for the former model has the same form of the Helmholtz equation for the latter. Because of this analogy, the theoretical results originally obtained in~\cite{izr01a} for a Kronig-Penney model with purely structural disorder immediately found experimental applications in the analysis of the transmission properties of single-mode waveguides~\cite{kuh00a,kro02, kuh08a}. It was therefore natural to ask whether the theoretical predictions concerning the effects of cross-correlations in the aperiodic Kronig-Penney model could find an experimental confirmation in suitable random waveguides. The main goal of this paper is to analyse this question, which we answer in the affirmative.

We considered microwave transmission of the first mode in a Q1D waveguide open on both ends and with no intermode coupling. The waveguide contained scatterers constituted by brass bars situated at random positions (structural disorder) and having random heights (compositional disorder).
We found that cross-correlations of the random spacings with the heights have a sizable effect on the transmission properties of the waveguide in
agreement with theoretical predictions. This is a remarkable finding, because the theoretical results for the localisation length are derived for an infinite array with weak disorder, whereas the experimental measures were obtained using a setup of finite size, with a modest number of scatterers (40 bars) and relatively strong disorder. The experimental data therefore show that the effect of disorder correlation is robust and is not washed away by other features of the apparatus (such as absorption) which are neglected in the simplified theoretical description.

This paper is organised as follows. In \sref{theo} we outline the theoretical framework. In \sref{exp} we describe the setup. Our experimental findings are presented in \sref{results}. We summarise our findings and expose our conclusions in \sref{conclu}.

\section{Theoretical considerations}
\label{theo}

We analyse the transmission of microwaves in a rectangular waveguide
with internal scatterers, constituted by brass bars situated at random
positions and having random heights.
If one assumes that the bars can be assimilated to point-like scatterers
and represented as delta-barriers, the propagation of a single mode in
the waveguide can be described using a Helmholtz equation which has the
same mathematical form of the Schr\"{o}dinger equation for the aperiodic
Kronig-Penney model with delta-barriers~\cite{kuh00a,kro02}
\begin{equation}
-\psi^{\prime \prime}(x) + \left[ \sum_{n=-\infty}^{\infty} (U + u_{n})
\delta(x - n - a_{n}) \right] \psi(x) = E \psi(x).
\label{kpmodel}
\end{equation}
For the sake of simplicity, here and in the rest of the paper we use
energy units such that $\hbar^{2}/2m = 1$ and set the lattice step as
unit length.

The model~\eref{kpmodel}, as the associated waveguide, has a twofold
randomness: on the one hand, the positions of the delta-barriers present
random shifts $a_{n}$ with respect to the lattice sites (structural
disorder). On the other hand, compositional disorder is introduced via
the fluctuations $u_{n}$ of the barrier strengths around the mean
value $U$.
The statistical properties of model~\eref{kpmodel} are defined by
assuming that the random variables $u_{n}$ and $a_{n}$ have vanishing
averages and known binary correlators. In the case of weak-disorder
no further specification of the statistical features of the model is
required.
The weak-disorder case is identified by the
conditions~\cite{her08,her10b,her10c}
\begin{displaymath}
\begin{array}{cccc}
\langle u_{n}^{2} \rangle \ll U^{2}, &
\langle \Delta_{n}^{2} \rangle E \ll 1, & \mbox{ and } &
\langle \Delta_{n}^{2} \rangle U \ll 1 ,
\end{array}
\end{displaymath}
with $\Delta_{n} = a_{n+1} - a_{n}$ representing the relative displacement of
two contiguous barriers.

If disorder is weak, one can obtain an analytical expression for the
localisation length $l_{loc}$ of the electronic states $\psi$ of
the model~\eref{kpmodel}~\cite{her08,her10b,her10c}.
Following the Hamiltonian map approach, one can show that, within the
second-order approximation, the electronic states have an inverse
localisation length equal to
\begin{equation}
\begin{array}{ccl}
l_{\mathrm{loc}}^{-1} & = & \displaystyle
\frac{1}{8 \sin^{2}\kappa}
\left[ \frac{\sin^{2} k}{k^{2}} \langle u_{n}^{2} \rangle
W_{1}(\kappa) + U^{2} \langle \Delta_{n}^{2} \rangle W_{2}(\kappa) \right. \\
& - & \displaystyle \left.
2 U \frac{\sin(k)}{k} \langle u_{n} \Delta_{n} \rangle
\cos \left( \kappa \right) W_{3}(\kappa) \right] .
\end{array}
\label{invloc}
\end{equation}
In \eref{invloc} the symbols $k$ and $\kappa$ represent respectively
the propagation wavevector within the wells $k = \sqrt{E}$ and the Bloch
wavevector $\kappa$ of the quantum particle.
The two quantities are linked by the condition
\begin{equation}
\cos \kappa = \cos k + \frac{U}{2k} \sin k ,
\label{kpband}
\end{equation}
which defines the band structure of the Kronig-Penney model~\eref{kpmodel}
in the absence of disorder.
The functions $W_{i}(\kappa)$ in \eref{invloc}, on the other hand,
are the Fourier transforms of the normalised binary correlators
of the random variables $u_{n}$ and $\Delta_{n}$, i.e.,
\begin{equation}
\begin{array}{ccl}
W_{1} \left( \kappa \right) & = & \displaystyle
1 + 2 \sum_{l=1}^{\infty} \frac{\langle u_{n}u_{n+l} \rangle}
{ \langle u_{n}^{2} \rangle} \cos(2 \kappa l) \\
W_{2} \left( \kappa \right) & = & \displaystyle
1 + 2 \sum_{l=1}^{\infty} \frac{\langle \Delta_{n}\Delta_{n+l} \rangle}
{\langle \Delta_{n}^{2} \rangle} \cos (2 \kappa l) \\
W_{3} \left( \kappa \right) & = & \displaystyle
1 + 2 \sum_{l=1}^{\infty} \frac{\langle u_{n}\Delta_{n+l} \rangle}
{\langle u_{n} \Delta_{n} \rangle} \cos (2 \kappa l) . \\
\end{array}
\label{powerspectra}
\end{equation}
\Eref{invloc} shows that the localisation length diverges
(within the second-order approximation) in any given energy interval
with vanishing power spectra \eref{powerspectra}.
Therefore an effective localisation-delocalisation transition
can occur in the 1D Kronig-Penney model \eref{kpmodel} provided that
the disorder exhibits the specific long-range correlations which make
the power spectra \eref{powerspectra} vanish over a continuous energy
range.

In order to obtain a specific localisation-delocalisation transition
with pre-assigned mobility edges, one must know how to construct two
random sequences $\{u_{n}\}$ and $\{\Delta_{n}\}$ such that the
corresponding power spectra \eref{powerspectra} vanish in pre-defined
energy intervals.
A solution to this problem was presented in~\cite{her10c}; the basic
idea is to generate two independent white-noise sequences which, after
having been cross-correlated, can be endowed with the appropriate
self-correlations with a filtering process.
More precisely, one starts with two sequences $\{ X_{n}^{(1)}\}$ and
$\{ X_{n}^{(2)}\}$ of independent identically distributed random variables.
Two cross-correlated white noises can then be obtained by using the transformation
\begin{displaymath}
\begin{array}{ccl}
Y_{n}^{(1)} & = & X_{n}^{(1)} \cos \eta + X_{n}^{(2)} \sin \eta \\
Y_{n}^{(2)} & = & X_{n}^{(1)} \sin \eta + X_{n}^{(2)} \cos \eta , \\
\end{array}
\end{displaymath}
where $\eta$ determines the degree of
cross-correlation of the $Y$ variables. The values of $\eta$ lie in
the interval $[-\pi/4, \pi/4]$ with $\eta = \pm \pi/4$ corresponding
to the extreme cases of total positive and negative cross-correlation,
while for $\eta = 0$ the cross-correlations vanish.
The cross-correlated white noises $Y_{n}^{(1)}$ and $Y_{n}^{(2)}$ can
then acquire the required self-correlations by means of the convolution
products
\begin{equation}
\begin{array}{ccl}
u_{n} & = & \displaystyle
\sum_{l=-\infty}^{\infty} \alpha_{l} Y_{n-l}^{(1)} \\
\Delta_{n} & = & \displaystyle
\sum_{l=-\infty}^{\infty} \beta_{l} Y_{n-l}^{(2)} \\
\end{array}
\label{convol}
\end{equation}
in which the coefficients $\alpha_{n}$ and $\beta_{n}$ are derived by
the pre-defined power spectra $W_{1}$ and $W_{2}$ via the equations
\begin{equation}
\begin{array}{ccl}
\alpha_{n} & = & \displaystyle
\frac{2}{\pi} \int_{0}^{\pi/2} \sqrt{\langle u_{l}^{2} \rangle W_{1}(x)}
\cos \left( 2 n x \right) \mathrm{d}x \\
\beta_{n} & = & \displaystyle
\frac{2}{\pi} \int_{0}^{\pi/2} \sqrt{\langle \Delta_{l}^{2} \rangle W_{2}(x)}
\cos \left( 2 n x \right) \mathrm{d}x .\\
\end{array}
\label{ab}
\end{equation}
After substituting the random variables defined by \eref{convol}
and \eref{ab} in \eref{invloc}, the inverse localisation length
takes the form
\begin{equation}
\begin{array}{ccl}
l_{\mathrm{loc}}^{-1} & = & \displaystyle
\frac{1}{8 \sin^{2} \kappa} \left[
\left( \frac{\sin k}{k} \right)^{2} \langle u_{n}^{2} \rangle W_{1}(\kappa) +
U^{2} \langle \Delta_{n}^{2} \rangle W_{2}(\kappa) \right. \\
& - & \displaystyle \left.
2 \left| \frac{\sin k}{k} \right| U \sqrt{\langle u_{n}^{2} \rangle
\langle \Delta_{n}^{2} \rangle W_{1}(\kappa) W_{2}(\kappa)} \;
\cos \kappa \sin \left( 2 \eta \right) \right] .
\end{array}
\label{ll}
\end{equation}

As a simple example of delocalisation transition produced by self-correlations
and modulated by cross-correlations, one can consider the case in which
both structural and compositional disorder have a power spectrum of the form
\begin{equation}
W_{1}(\kappa) = W_{2}(\kappa) = \left\{ \begin{array}{ccl}
\displaystyle
\frac{\pi}{2a \left(\kappa_{2} - \kappa_{1}\right)} &
\mbox{ if } & \kappa \in \left[ \kappa_{1}, \kappa_{2} \right] \\
0 & \mbox{ if } &
\kappa \in \left[ 0, \kappa_{1} \right] \cup
\left[ \kappa_{2}, \frac{\pi}{2}\right] \\
\end{array} \right.
\label{cond1}
\end{equation}
with $0 < \kappa_{1} < \kappa_{2} < \pi/2$.
Note that it is enough to consider the $[0,\pi/2]$ interval in the
previous definition because the power spectra \eref{powerspectra} are
even functions of period $\pi$.
The binary self-correlators corresponding to the power spectra \eref{cond1}
decrease with a power law
\begin{equation}
\frac{\langle u_{l+n} u_{l} \rangle}{\langle u_{l}^{2} \rangle} =
\frac{\langle \Delta_{l+n} \Delta_{l} \rangle}{\langle \Delta_{l}^{2} \rangle}
 = \frac{1}{2 \left(\kappa_{2} - \kappa_{1} \right)n}
\left[ \sin \left( 2 \kappa_{2} n \right) -
\sin \left( 2 \kappa_{1} n \right) \right] ,
\label{selfcor}
\end{equation}
which shows that the random variables $u_{n}$ and $\Delta_{n}$ have
long-range self-correlations.
To simplify further the physical picture, one can consider the case in
which structural and compositional disorder have the same strength:
\begin{equation}
\langle u_{n}^{2} \rangle = \langle \Delta_{n}^{2} \rangle = \sigma^{2} .
\label{cond2}
\end{equation}
When conditions \eref{cond1} and \eref{cond2} are met, the
inverse localisation length \eref{ll} for $\kappa > 0$ reduces to the form
\begin{equation}\fl
l_{\mathrm{loc}}^{-1} =
\left\{ \begin{array}{cl}
\displaystyle
\frac{\pi \sigma^{2}}{16 (\kappa_{2} - \kappa_{1}) \sin^{2}\kappa}
\left[ \left( \frac{\sin k}{k} \right)^{2} + U^{2} -
2U \left| \frac{\sin k}{k} \right| \cos \kappa \sin (2 \eta) \right]
&, \mbox{if } \kappa \in I \\
0
&, \mbox{otherwise} \\
\end{array}\right.
\label{ll2}
\end{equation}
with $I = [\kappa_{1},\kappa_{2}] \cup [\pi - \kappa_{2}, \pi - \kappa_{1}]$.
In this case the electronic states are expected to be localised in the
intervals $[\kappa_{1}, \kappa_{2}]$ and $[\pi - \kappa_{2}, \pi - \kappa_{1}]$
and extended (within the second-order approximation) in the complementary
regions of the $[0, \pi]$ interval (which is the left half of
the first Brillouin zone).

\begin{figure}
\centerline{\includegraphics[]{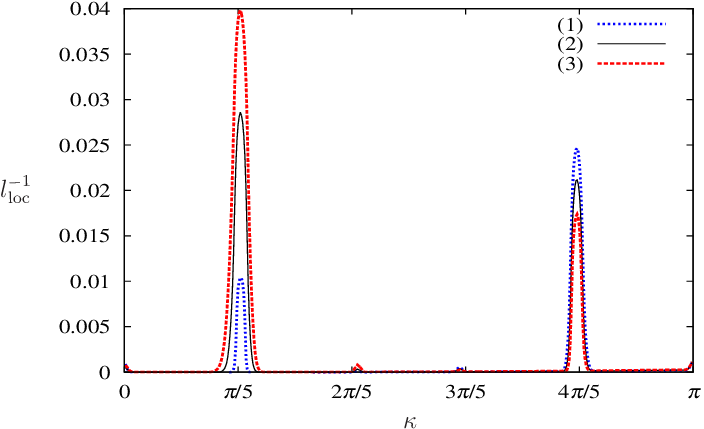}}
\caption{\label{lyapexp} (Colour online)
Inverse localisation length $l_{\mathrm{loc}}^{-1}$ versus Bloch vector $\kappa$.
The solid black line (2) represents the results obtained in the absence of
cross-correlations between structural and compositional disorder, while the
blue dotted line (1) and the red dashed line (3) respectively correspond to
the cases of total positive and negative cross-correlations.}
\end{figure}

A numerical evaluation of the inverse localisation length confirms
this conclusion, as can be seen from \fref{lyapexp}.
The data represented in \fref{lyapexp} were obtained for
$\kappa_{1} = 0.20 \pi$ and $\kappa_{2} = 0.21 \pi$, while the mean field and the
disorder strength were set equal to $U = 1$ and
$\sqrt{\langle u_{n}^{2} \rangle} = \sqrt{\langle \Delta_{n}^{2} \rangle} = 0.05$.
The long-range self-correlations \eref{selfcor} enhance the
localisation of the electronic states in the two narrow regions
$[0.20 \pi, 0.21 \pi]$ and $[0.79 \pi, 0.80 \pi]$ and delocalise
the other states.
Cross-correlations, on the other hand, modulate the
localisation-delocalisation transition, making it more or less
pronounced according to the sign of the $\cos \kappa \sin(2\eta)$ term
in the right-hand side of \eref{ll2}.

\begin{figure}
\centerline{\includegraphics[]{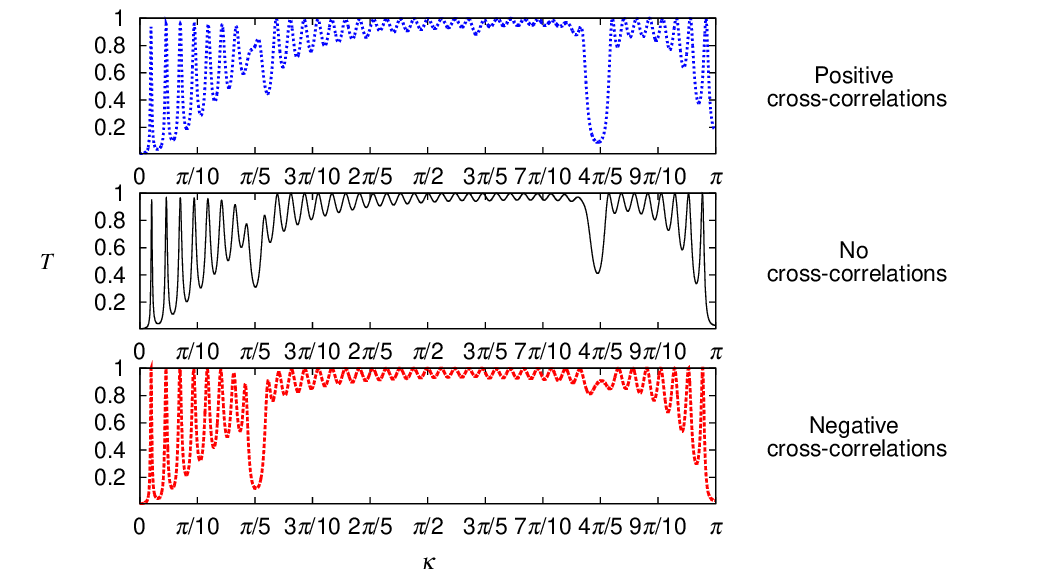}}
\caption{\label{theortrans} (Colour online) Transmission coefficient $T$
versus Bloch wavevector $\kappa$ in an aperiodic Kronig-Penney model of $N=40$
sites. The cases of total positive and negative cross-correlations
are compared with the case in which cross-correlations are absent.
The represented data were obtained for the same parameters of
the model used in the case of \fref{lyapexp}, i.e., $U = 1$ and
$\sqrt{\langle u_{n}^{2} \rangle} = \sqrt{\langle \Delta_{n}^{2} \rangle} = 0.05$.}
\end{figure}

Obviously, the localisation-delocalisation transition produced by
long-range self-correlations has repercussions on the transmission
properties of a finite aperiodic Kronig-Penney model sandwiched between
two perfect leads.
When the longitudinal size of the random sample is larger than the
localisation length of the localised states, it is natural to expect
that a continuum of delocalised states manifests itself in the form of a
transparent energy window, while the regions of enhanced localisation
become low-transmittivity gaps.
As for cross-correlations, they can either enhance or reduce the opacity
of the low-transmittivity intervals.
Numerical studies confirmed this expectations~\cite{her10c}.
They also showed that the effects of self- and cross-correlations can be
detected in random samples of moderate size, as can be seen in \fref{theortrans},
which represents the transmission coefficient $T$ of
a random Kronig-Penney model of $N=40$ sites for a specific realisation of
the disorder.

One can see that self-correlations create transmission gaps corresponding
to the windows of localised states.
Positive cross-correlations increase the transmittivity in the low-energy
gap (almost removing the gap itself) and further reduce it in the high-energy
gap, whereas for negative cross-correlations the inverse effect is observed
in the two gaps.
As a final comment on this issue we note that, when one considers
transmission in {\em short} random samples, sample-to-sample
fluctuations may prevent the transmission coefficient from following
closely the behaviour of the theoretical localisation length for
some disorder realisations.
In the next section we will discuss how the effects of cross-correlations
appear in our random waveguide.

\section{Experimental setup and measurement technique}
\label{exp}

We describe the experimental setup and the measurement technique in three
steps. We first present the apparatus and explain how we measured the
mode-resolved microwave transport through the waveguide. Then we
explain how we realised the random cross-correlated lattice inside the
waveguide. Finally we discuss how the experimental band structure was
related to that of the theoretical model \eref{kpmodel} thereby
establishing a functional relation between the wavevector $k_{x}$ and
the Bloch vector $\kappa$.

\subsection{Microwave measurement}
\label{mm}

\begin{figure}
\centerline{\includegraphics[width=0.9\textwidth]{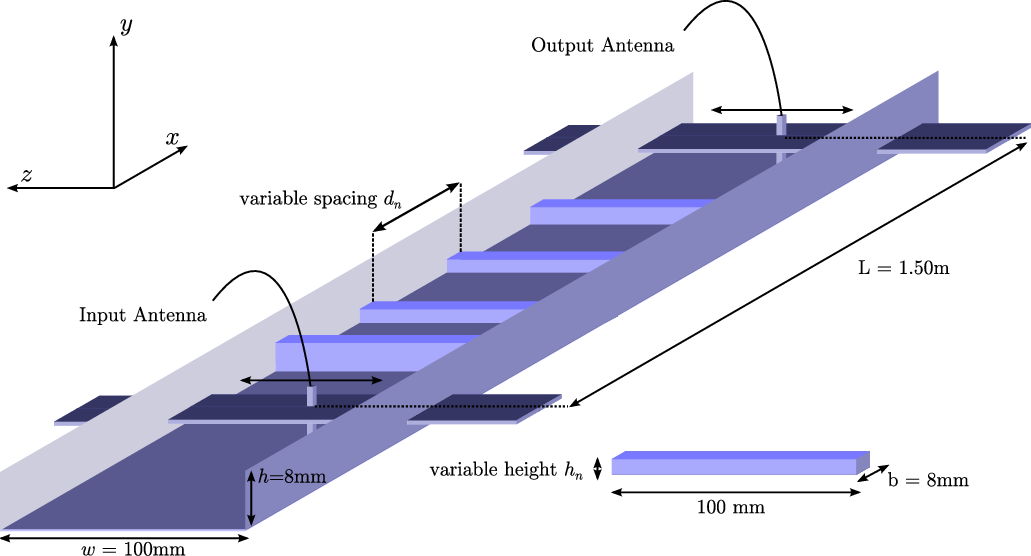}}
\caption{\label{setup} (Colour online)
Schematic picture of the experimental setup with top plate removed (not to
scale). Brass bar inlays can be freely placed inside the waveguide. The
number of bars varies according to the spacing $d_{n}$. Antennas shown here
with parts of the top plate can be moved separately along the $z$-direction.
The distance along the $x$-axis between the antennas is $L=1.50$\,m,
the total waveguide length is 2.38\,m. Absorbers at the
ends of the waveguide are omitted.}
\end{figure}

For our experiment we used a variant of the setup employed in~\cite{die11a}
to analyse transport in quasi-1D structures.
The setup consists of a rectangular waveguide. The top plate can be lifted
so that scatterers can be placed at will inside the waveguide.
\Fref{setup} shows the waveguide with the top plate removed and four
bars placed inside as an example of scattering structure.

Two antennas are set at positions $x_{\mathrm{in}}=0$ and
$x_{\mathrm{out}}=1.5 \mbox{ m}$, with one serving as an emitter and
the other as a receiver. The distance between the antennas,
$L = x_{\mathrm{out}} - x_{\mathrm{in}} = 1.5 \mbox{ m}$ represents the effective
length of the waveguide, whose width and height are $w = 100 \mbox{ mm}$
and $h = 8 \mbox{ mm}$.
The antennas are plugged into the waveguide through mobile slides so
that they can be moved freely from $0$ to $w$ along the transversal
$z$-direction.
Microwave absorbers are placed between each antenna and the corresponding
end of the waveguide to avoid back-reflection from the open end.
The total length of the apparatus is $2.38 \mbox{ m}$.

A microwave vector network analyser allowed us to measure the transmission
coefficient of the waveguide from the input antenna at
$(x_{\mathrm{in}},z_{\mathrm{in}})$ to the output antenna at
$(x_{\mathrm{out}},z_{\mathrm{out}})$, i.e., the scattering matrix
$S(x_{\mathrm{in}}, z_{\mathrm{in}}; x_{\mathrm{out}}, z_{\mathrm{out}})$.
Because the antennas are fixed in the $x$-direction, in what follows we
write $S(z_{\mathrm{in}},z_{\mathrm{out}})$ as a short-hand notation for
$S(0,z_{\mathrm{in}};L,z_{\mathrm{out}})$.
We measured the waveguide transmission coefficient as a function of
the frequency $\nu$ for the lowest-frequency TE componentlowest-frequency
transverse electric (TE) component of the electromagnetic field.
In fact, for a given frequency $\nu$ there are several propagating waves
with wavenumber $|k| = 2 \pi \nu /c$.
Due to the confinement along the width of the waveguide, the
$z$-component of the wavevector takes discrete values, $k_{z}^{(n)} = \pi n/w$,
corresponding to propagating modes with longitudinal wavenumbers
\begin{displaymath}
k_{x}^{(n)} = \sqrt{k^{2} - ( \pi n/w)^{2}}
\end{displaymath}
with $n = 1, 2, \ldots, N_{w}$.
The total number $N_{w}$ of propagating modes is determined by the
integer part $[[ \cdots ]]$ of the mode parameter $kw/\pi$:
\begin{displaymath}
N_{w} = [[ kw/\pi]] .
\end{displaymath}

To separate the different modes we measure $S(z_{\mathrm{in}},z_{\mathrm{out}})$
at several positions $z_{\mathrm{in}}$ and $z_{\mathrm{out}}$ distributed over
the whole width $w$.
The transmission between the initial point $(x=0,y_{1})$ and the final point
$(x=L,y_{2})$ can be decomposed into all contributing modes
\begin{equation}
S(z_\mathrm{in},z_\mathrm{out}) = \sum_{nm} S_{nm} \sin\left(\frac{m\pi
z_\mathrm{in}}{w}\right) \sin\left(\frac{n\pi z_\mathrm{out}}{w}\right).
\label{streuung}
\end{equation}
From \eref{streuung} one can obtain the scattering matrix
in mode representation with a twofold sine transform. The matrix
element $S_{nm}$ denotes the scattering from the $m$-th to the $n$-th mode.

If all propagating modes were considered, the waveguide would correspond
to a quasi-1D open strip~\cite{die11a}, while the transmission coefficient
for the first mode would be related to the scattering matrix via the identity
\begin{displaymath}
T_{1} = \sum_{n=1}^{N_{w}} |S_{1n}|^{2} .
\end{displaymath}
However, we found that, as long as the scatterers fill homogenously the
waveguide along the $z$ direction, the individual modes propagate independently
of each other through the waveguide. Therefore we chose bars whose length
perfectly matched the waveguide width keeping a translation invariance in the $y$ direction.
In this way, as in~\cite{die11a}, we obtained a waveguide in which
the mode-mode couplings were usually negligible.
This made the scattering matrix $S_{mn}$ practically diagonal (so that
$T_{1} \simeq |S_{11}|^{2}$).
The absence of intermodal scattering allowed us to restrict our
measurements to the first propagating mode: we were thus able
to put the experimental setup in correspondence with the strictly 1D
theoretical model \eref{kpmodel}.
Note that the equivalence of the waveguide to a 1D device did not require
that the setup itself be strictly 1D.

\subsection{Correlated random scattering structures}
\label{scastr}

We used brass bars as scatterers in the waveguide. As explained
in \sref{mm}, the requirement that the setup should behave as a 1D
device defined the length of the scattering bars.
The bar width $b=8 \mbox{ mm}$ was chosen as in~\cite{die11a}, because
it is a good trade-off between small width and sufficient scattering
strength.
The width of the bars and the finite length of the waveguide limited
the number of bars that we could insert: we formed a scattering array of
40 bars in total.

In contrast to~\cite{die11a}, in the present case the bars were not only
separated by random spacings $d_{n}$ (structural disorder), but also had
random heights $h_{n}$ (compositional disorder).
The bar heights spanned the range from $0.1 \mbox{ mm}$ to $0.8 \mbox{ mm}$
in steps of $0.1 \mbox{ mm}$.
We derived sequences of bar heights $h_{n}$ and spacings $d_{n}$ by scaling
the corresponding variables $u_{n}$ and $\Delta_{n}$ of the Kronig-Penney
model \eref{kpmodel} which we used to obtain the transmission coefficient
represented in \fref{theortrans} (see \sref{theo}).
In this way we obtained two random successions $\{h_{n}\}$ and $\{d_{n}\}$
with self-correlations of the form \eref{selfcor} with
$\kappa_{1} = 0.20 \pi$ and $\kappa_{2} = 0.21 \pi$.
This choice implied a rather high value of the transmittivity for most
frequencies, with a sharp drop in the two narrow gaps corresponding to
the intervals $[0.20 \pi, 0.21 \pi]$ and $[0.79 \pi, 0.80 \pi]$ of the
left half of the first Brillouin zone.
As in the example discussed in \sref{theo}, in addition to
long-range self-correlations, the random sequences of bar heights
and spacings were cross-correlated: we considered the cases of maximally
positive ($\eta = \pi/4$) and negative ($\eta = - \pi/4$)
cross-correlations as well as the case in which heights and spacings of
the bars were not cross-correlated ($\eta = 0$).

\begin{figure}
\centerline{\includegraphics{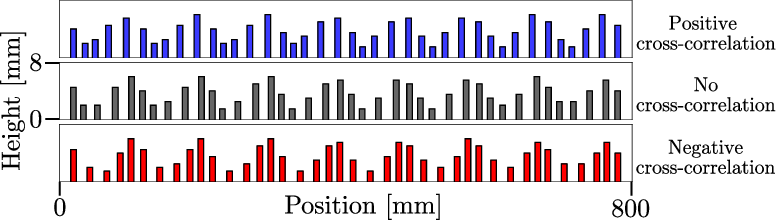}}
\caption{\label{bars} (Colour online)
Profile of the scattering arrangement within the waveguide for the
cases of positive (top), absent (middle), and negative (bottom)
cross-correlations.}
\end{figure}

In \fref{bars} we show the arrangements of the scattering bars within
the waveguide for the three cases discussed. The influence of the
cross-correlations is clearly visible. In fact, positive cross-correlations
tend to produce large spacings between high bars, while negative
cross-correlation reduce the distance separating high bars.

\subsection{Transmission bands and Bloch vector calculation}

\begin{figure}
\centerline{\includegraphics[]{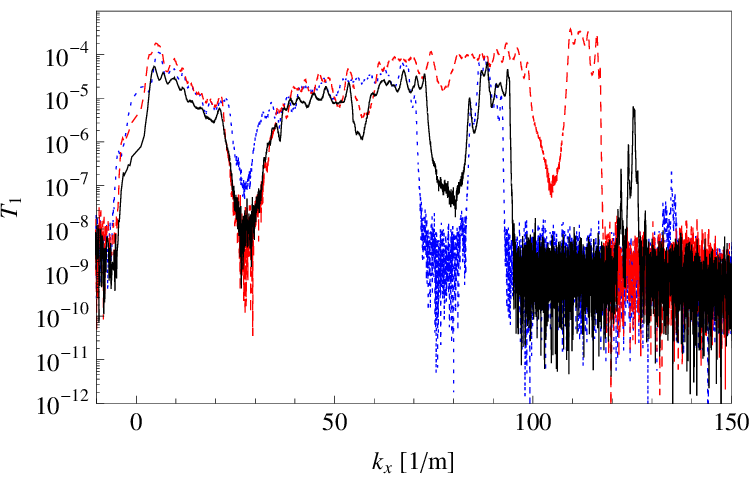}}
\caption{\label{rawband} (Colour online)
Transmission coefficient $T_{1}$ for the first propagating mode versus wave
vector $k_{x}$. The dotted line (blue) and the dashed line (red) respectively
correspond to the cases of maximal positive and negative cross-correlations.
The solid line (black) corresponds to the case in which no cross-correlations
are present.}
\end{figure}

The measured transmission for the first mode versus the longitudinal
wavevector is shown in \fref{rawband}.
One can clearly see the first band. The band edges are the same for the
cases of positive and absent cross-correlations, but in the case of
negative cross-correlations the band is broader and the second gap is
shifted towards higher wavenumbers $k_{x}$.
Note that in all the measurements we used the same bars, which were
only shuffled and put in different positions in the three cases.
This means that, although the constituents of the scattering arrangements
are the same, cross-correlations can broaden the transmission band.
We attribute this effect to a shadowing effect. In the case of negative
cross-correlations, in fact, high bars are separated by much smaller distances
than in the other cases.
Therefore the waves do not ``feel'' the same underlying lattice structure in
the case of negative cross-correlations and this is probably the origin of
the enlarged band.

To analyse the changes in the transmission gaps brought about by
cross-correlations, we need to know the band structure of the
experimental setup or, equivalently, the relation between the longitudinal
wavevector $k_{x}$ and the Bloch vector $\kappa$.

We followed an empirical approach to establish a correspondence
between $k_{x}$ and $\kappa$.
Specifically, we associated the edges of the transmission band to the
values $\kappa=0$ and $\kappa=\pi$ of the Bloch vector. We then used a
linear interpolation to extend the $k_{x}-\kappa$ correspondence to the
whole Brilllouin zone.
This corresponds to a linear fit $\kappa=a k$, which is consistent with
the fact that, as shown in~\cite{die11a}, the finite width of the bars can
be theoretically described by introducing an effective refractive index $a$
so that $k \rightarrow a k$.

\section{Experimental results}
\label{results}

\begin{figure}
\centerline{\includegraphics[]{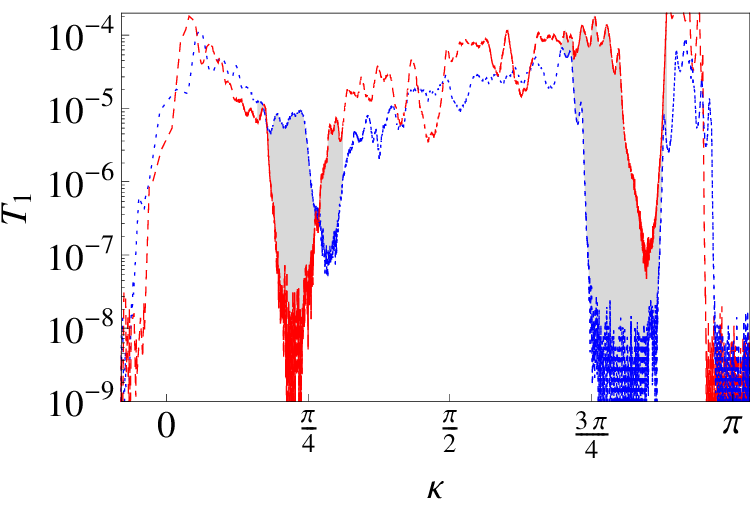}}
\caption{\label{posneg} (Colour online)
Transmission coefficient $T_{1}$ for the first propagating mode versus Bloch
wavevector $\kappa$. The dotted line (blue) and the dashed line (red)
respectively correspond to the cases of maximal positive and negative
cross-correlations. The shaded area highlights the variation of $T_{1}$
due to cross-correlations.}
\end{figure}

\begin{figure}
\centerline{\includegraphics[width=5in]{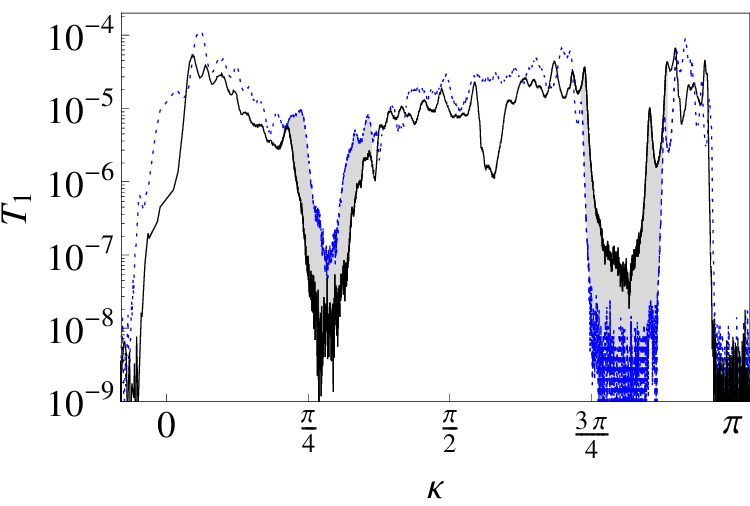}}
\caption{\label{zeropos} (Colour online)
Transmission coefficient $T_{1}$ for the first propagating mode versus Bloch
wavevector $\kappa$. The dotted line (blue) and the solid line (black)
respectively correspond to the cases of positive and absent
cross-correlations. The shaded area highlights the variation of $T_{1}$
due to positive cross-correlations.}
\end{figure}

\begin{figure}
\centerline{\includegraphics[width=5in]{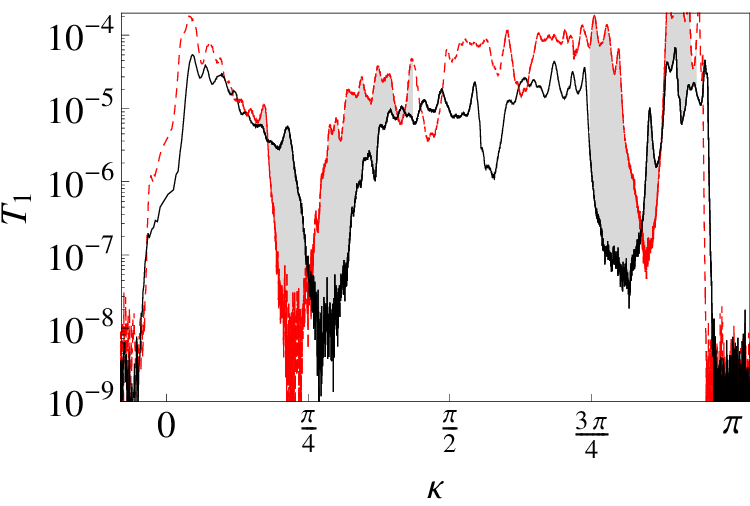}}
\caption{\label{zeroneg} (Colour online)
Transmission coefficient $T_{1}$ for the first propagating mode versus Bloch
wavevector $\kappa$. The dashed line (red) and the solid line (black)
respectively correspond to the cases of negative and absent
cross-correlations. The shaded area highlights the variation of
$T_{1}$ due to negative cross-correlations.}
\end{figure}

The theoretical results for the Kronig-Penney model, represented by the
inverse localisation length \eref{ll2} and the transmission coefficient
pictured in \fref{theortrans}, suggest that the transmission
gap in the right neighbourhood of $\kappa = \pi/5$ should be reduced or
enhanced according to whether the height-position correlations are
positive or negative, whereas the contrary should happen for the gap in
the left neighbourhood of $\kappa = 4 \pi/5$.
These expectations are confirmed by the experimental results, represented
in \fref{posneg}, \fref{zeropos}, and \fref{zeroneg}.

In \fref{posneg} we show the transmission coefficient $T_{1}$ for the
first propagating mode in the two extreme cases of maximum positive and
negative cross-correlations between compositional and structural disorders.
\Fref{zeropos} and \fref{zeroneg} respectively compare the transmission
coefficient $T_{1}$ in the absence of cross-correlations ($\eta = 0$)
with the same quantity measured when cross-correlations take the
maximal positive ($\eta = \pi/4$) and negative ($\eta = -\pi/4$) values.
The experimental data show that the transmission coefficient in the
opaque gaps is reduced or enhanced according to the theoretical
expectations.
This is made particularly evident by the comparison of \fref{theortrans}
and \fref{posneg}. In both cases switching from positive to negative
cross-correlations decreases the transmittivity in the first opaque
window and enhances it in the second gap.
\Fref{zeropos} and \fref{zeroneg}, on the other hand, agree well
with the behaviour of the inverse localisation length in the Kronig-Penney
model represented in \fref{lyapexp}: one can easily see, for instance,
that positive cross-correlations decrease localisation in the first
region of localised states and enhance it in the second interval
while, correspondingly, they make the first transmission gap more
shallow and deepen the second one. Negative cross-correlations have
the same effects, but with reversed gaps.
The shift to lower (higher) frequency of the induced lower (higher) gaps in case of negative cross-correlations seen in \fref{zeroneg} and in \fref{posneg} is due to the fact that the linear rescaling using the band edge is only an approximation.

\section{Conclusions}
\label{conclu}

This work is the first experimental study of the effects that
cross-correlations between different random potentials have on the
transport properties of a 1D waveguide.
We analysed the microwave transmission of the first mode
in a multimode waveguide with compositional and structural disorder where no mode coupling was present.
Specifically, we inserted brass
bars in the waveguide which acted as scatterers and were characterised
by self- and cross-correlated random heights and positions.
We showed how cross-correlations between the random heights and spacings
of the scatterers can modulate the transmittivity in the non-transparent
frequency windows created by long-range self-correlations of the disorder.
The results agree with the theoretical predictions previously obtained for
an aperiodic Kronig-Penney model.

The agreement between theoretical predictions and experimental data
is far from trivial, because the theoretical analysis was based on a
number of simplifying assumptions, such as weak disorder and
irrelevance of some experimental imperfections, e.g.\ absorption,
which are present in any actual waveguide. Besides, the theoretical
results for the transmission coefficient did not guarantee that the
effect of cross-correlations would be visible also in 1D devices of reduced
length.
The results obtained in the present experiments, therefore, support the
validity of the theoretical predictions and show that the effects of
cross-correlations are not eliminated by the reduced length of the
random waveguide.

\ack
L.T. gratefully acknowledges the support of CONACyT grant No. 150484.
Support by the DFG within the research group 760 ``Scattering Systems
with Complex Dynamics'' is acknowledged by O.D. and U.K.

\end{document}